\documentclass[a4paper]{jpconf}
\usepackage{graphicx}

\usepackage[square,sort&compress]{natbib}

\begin{document}
\title{Multiple synchrotron self-Compton modeling of gamma-ray flares in 3C~279}

\author{Marc T\"urler$^1$ and Claes-Ingvar Bj\"ornsson$^2$}

\address{$^1$ ISDC, Geneva Observatory, University of Geneva, ch. d'Ecogia 16, 1290 Versoix, Switzerland}
\address{$^2$ Department of Astronomy, Alba Nova University Center, Stockholm University, SE--106 91 Stockholm, Sweden}

\ead{marc.turler@unige.ch}

\begin{abstract}
The correlation often observed in blazars between optical-to-radio outbursts and gamma-ray flares suggests that the high-energy emission region shall be co-spatial with the radio knots, several parsecs away from the central engine. This would prevent the important contribution at high-energies from the Compton scattering of seed photons from the accretion disk and the broad-line region that is generally used to model the spectral energy distribution of low-frequency peaking blazars. While a pure synchrotron self-Compton model has so far failed to explain the observed gamma-ray emission of a flat spectrum radio quasar like 3C~279, the inclusion of the effect of multiple inverse-Compton scattering might solve the apparent paradox. Here, we present for the first time a physical, self-consistent SSC modeling of a series of shock-waves in the jet of 3C~279. We show that the analytic description of the high-energy emission from multiple inverse-Compton scatterings in the Klein-Nishina limit can fairly well account for the observed gamma-ray spectrum of 3C 279 in flaring states.
\end{abstract}

\section{Introduction}
\label{introduction}
The radio-to-infrared emission of blazars is well understood as being synchrotron emission associated to bright structures observed to propagate --- often with apparent superluminal motion --- in a relativistic jet at parsec-scale distances \cite{SWV02}. The preferred model for the emission of these knots are shock waves resulting from disturbances at the base of the jet that move outward, become supersonic, compress the gas locally, and accelerate particles at the shock front \cite{MG85}. In order to test the validity of this shock-in-jet model, a tool was developed to fit the multi-frequency light-curves of the bright quasar 3C~273 with a series of model outbursts \cite{TCP99,TCP00}. The same approach was then also used to describe the flaring behavior of other sources: the blazar 3C~279 \cite{LTV06}, and the micro-quasars GRS~1915+105 \cite{TCC04} and Cyg~X-3 \cite{LTH07,MRT09}. While these studies still relied on model parameters that are observables, we now developed a fully physical parametrization of the synchrotron emission of shock waves \cite{T11}.

The correlation often observed in blazars between optical-to-radio outbursts and gamma-ray flares suggests that the high-energy emission is closely linked to the synchrotron component (e.g. \cite{LVT11}). If the emission arises parsecs away from the central engine, the seed photons for inverse-Compton scattering cannot be of external origin. However, for a low-energy peaking blazar like 3C 279, it has been shown that such a synchrotron self-Compton (SSC) scenario is not able to account for the observed gamma-ray emission \cite{HBA01,LVT05}. While this is true for single scatterings, the effect of multiple-inverse Compton (MIC) orders has been reconsidered recently to solve this apparent paradox \cite{B10}. We present here a first attempt to apply an analytical model for multiple synchrotron self-Compton (MSSC) emission of shock waves to a broad dataset of 3C~279.

\section{Theory}
\label{theory}
This section aims at linking the theoretical model for MIC scattering \cite{BA00,B10} with the parameters and the formalism used for the synchrotron emission of shock waves in a relativistic jet \cite{T11}. As a starting point we link the optical depth $\tau_0$ of \cite{B10} with the normalization $K$ of the electron energy distribution, assumed to be of the form $n_{\mathrm{e}}(\gamma)=K\,\gamma^{-p}$ for $\gamma_{\mathrm{min}}\!\le\!\gamma\!\le\!\gamma_{\mathrm{max}}$, where $n_{\mathrm{e}}$ is the electron number density and $\gamma$ is the Lorentz factor of the relativistic electrons. For $p>2$ and $\gamma_{\mathrm{max}}\gg \gamma_{\mathrm{min}}$, integrating  $n_{\mathrm{e}}(\gamma)$ leads to $K=(p-1)\gamma_{\mathrm{min}}^{p-1}\,n_{\mathrm{e}}$ such that Eq.~(7) of \cite{B10} becomes:
 \begin{equation}
\label{tau0}
\tau_0 = \frac{8\,\sigma_{\mathrm{T}}}{3(p-2)}\,x_R\,K\,,
\end{equation}
where $\sigma_{\mathrm{T}}$ is the Thomson cross-section and $x_R = f_R\,R$ is the thickness of the emitting slab containing the shocked plasma behind the shock front, which is assumed to be a constant fraction $f_R$ of the jet radius $R$ \cite{T11}.

For a given electron energy distribution $n_{\mathrm{e}}(\gamma)$, the MIC spectrum is determined by two main parameters $\alpha$ and $\gamma_{\mathrm{cool}}$. The spectral index $\alpha$ ($F_{\nu}\propto \nu^{-\alpha}$) connects the $\nu\,F_{\nu}$ maxima of the synchrotron and the inverse-Compton components. The Lorentz factor $\gamma_{\mathrm{cool}}$ corresponds to the energy of electrons that have just enough time to cool down while crossing the emission region, i.e. $x(\gamma_{\mathrm{cool}}) = \beta_{\mathrm{rel}} c t_{\mathrm{cool}}(\gamma_{\mathrm{cool}}) \equiv x_R$, where $\beta_{\mathrm{rel}} c$ is the average speed of electrons along the jet relative to the shock front.
In the MIC scenario, there are three distinct cases \cite{B10}:
\begin{eqnarray}
\mbox{A)} & \gamma_{\mathrm{cool}}>\gamma_{\mathrm{min}} \mbox{ and } \alpha>(p-1)/2 & \Longrightarrow \tau_0=\gamma_{\mathrm{cool}}^{\,p-1-2\alpha}<1\\
\mbox{B)} & \gamma_{\mathrm{cool}}>\gamma_{\mathrm{min}} \mbox{ and } \alpha<(p-1)/2 & \Longrightarrow \tau_0=\gamma_{\mathrm{min}}^{\,p-1-2\alpha}>1\\
\mbox{C)} & \gamma_{\mathrm{cool}}<\gamma_{\mathrm{min}} & \Longrightarrow \tau_0=\gamma_{\mathrm{min}}^{p-1-2\alpha}\left(\frac{\gamma_{\mathrm{min}}}{\gamma_{\mathrm{cool}}}\right)^{\alpha}.
\end{eqnarray}
In cases A and B,  $\alpha=(p-1)/2$ coincides with $\tau_0=1$ and we note the corresponding value of $\gamma_{\mathrm{cool}}$ as $\gamma_{\mathrm{cool},0}$.
One can derive a simple equation linking $\gamma_{\mathrm{cool}}$ to $\alpha$ as:
\begin{equation}
\label{eq1}
\left(\frac{\gamma_{\mathrm{cool}}}{y}\right)^{\alpha} = \left(\frac{\gamma_{\mathrm{cool},0}}{y}\right)^{(p-1)/2},
\end{equation}
where $y\equiv mc^2/(h\nu_B)$ and $\nu_B\equiv eB/(2\pi mc)$ is the cyclotron frequency for an electron mass, $m$, and charge, $e$, in a magnetic field, $B$. Eq.~(\ref{eq1}) is valid in case A and also in case C, if in the latter the label ``0'' denotes $\alpha = (p-1)/2$ rather than $\tau_0=1$. For case B, there is an additional term $(\gamma_{\mathrm{min}}/\gamma_{\mathrm{abs}})^{p-1-2\alpha}$ on the left-hand side (LHS). This is due to the fact that in this case the Compton-scattered flux is dominated by the lowest energy synchrotron photons, i.e. those radiating at the synchrotron self-absorption frequency $\nu_{\mathrm{abs}}=\nu_B\gamma_{\mathrm{abs}}^2$.

In Eq.~(\ref{eq1}) we implicitly assumed that $\gamma_{\mathrm{abs}}<\gamma_{\mathrm{cool}}$. When $\gamma_{\mathrm{abs}}>\gamma_{\mathrm{cool}}$, the corresponding equation is:
\begin{equation}
\label{eq2}
\,\left(\frac{\gamma_{\mathrm{cool}}}{\gamma_{\mathrm{abs}}}\right)^{p-1-2\alpha}\left(\frac{\gamma_{\mathrm{cool}}^2}{y\,\gamma_{\mathrm{abs}}}\right)^{\alpha}=\left(\frac{\gamma_{\mathrm{cool},0}^2}{y\,\gamma_{\mathrm{abs},0}}\right)^{(p-1)/2}\,,
\end{equation}
where $\gamma_{\mathrm{abs},0}$ is the value of $\gamma_{\mathrm{abs}}$ when $\tau_0=1$ and the equation reduces to Eq.~(\ref{eq1}) when $\gamma_{\mathrm{abs}} \rightarrow \gamma_{\mathrm{cool}}$. For case B, a modification is again needed to the LHS of Eq.~(\ref{eq2}): the ratio $\gamma_{\mathrm{cool}}/\gamma_{\mathrm{abs}}$ has to be replaced by $\gamma_{\mathrm{min}}/\gamma_{\mathrm{abs}}$.
From Eqs.~(13) and (18) of  \cite{B10}, one notes that  the right-hand side (RHS) of both Eqs.~(\ref{eq1}) and (\ref{eq2}) is given by:
\begin{equation}
\label{eq3}
\mbox{RHS}=\frac{\gamma_{\mathrm{min}}^{p-2}}{y}\,\frac{2\beta_{\mathrm{rel}}}{\tau_0}\,\frac{U_{\mathrm{e}}}{U_B} = \frac{6\pi\,mc^2\,\beta_{\mathrm{rel}}}{\sigma_{\mathrm{T}}\,y\,x_R\,B^2}=\frac{3\,h\,e\,\beta_{\mathrm{rel}}}{mc\,\sigma_{\mathrm{T}}\,x_R\,B}\,,
\end{equation}
where $U_{\mathrm{e}}=K\,mc^2\,\gamma_{\mathrm{min}}^{2-p}/(p-2)$ and $U_B=B^2/(8\pi)$  are the energy densities of the electrons and the magnetic field, respectively,  and where we used Eq.~(\ref{tau0}) for $\tau_0$. This equation is to be calculated for a given jet model. We now have to obtain equations for $\alpha$ that are independent of $\gamma_{\mathrm{cool}}$ in each case A, B and C, and both for 1) $\gamma_{\mathrm{abs}}<\gamma_{\mathrm{cool}}$ with Eq.~(\ref{eq1}) and for 2) $\gamma_{\mathrm{abs}}>\gamma_{\mathrm{cool}}$ with Eq.~(\ref{eq2}).
We get the following equations for $\alpha$:
\begin{eqnarray}
\mbox{A1)} & [2\ln{y}]\,\alpha^2+[\,\ln{\tau_0}+2\ln{(\mbox{RHS})}\!-\!(p\!-\!1)\ln{y}\,]\,\alpha-(p\!-\!1)\ln{(\mbox{RHS})}=0\\
\mbox{A2)} & [2\ln{(y/\gamma_{\mathrm{abs}})}]\,\alpha^2+[(p\!-\!1)\ln{(\gamma_{\mathrm{abs}}^3/y)}+2\ln{(\mbox{RHS})}]\,\alpha-(p\!-\!1)\ln{(\mbox{RHS}\,\gamma_{\mathrm{abs}}^{p-1}/\tau_0)}=0\\
\mbox{B)} & \alpha=(p\!-\!1)/2 - \ln{\tau_0}/(2\ln{\gamma_{\mathrm{min}}})\\
\mbox{C1)} & \alpha=\ln{(\gamma_{\mathrm{min}}^{p-1}/(\mbox{RHS}\,\tau_0))}/\ln{(y\,\gamma_{\mathrm{min}})}\\
\mbox{C2)} & [\ln{(y/\gamma_{\mathrm{abs}})}]\,\alpha^2+[(p\!-\!1)\ln{(\gamma_{\mathrm{min}}\gamma_{\mathrm{abs}})}+\ln{(\mbox{RHS})}]\,\alpha-(p\!-\!1)\ln{(\gamma_{\mathrm{min}}^{p-1}/\tau_0)}=0
\end{eqnarray}
Pratically, as $\gamma_{\mathrm{abs}}$ is not know a priori, we first derive $\alpha$ for the cases A1, B and C1, which are independent of $\gamma_{\mathrm{abs}}$. We then calculate $\gamma_{\mathrm{cool}}$ for each of the three cases A1, B1 and C1 using Eq.~\ref{eq1} and then $\gamma_{\mathrm{abs}}$ given by:
\begin{equation}
\label{g_abs}
\gamma_{\mathrm{abs}}^{p+4+\alpha}=\frac{e^2 g_{\kappa}(p)}{16\,mc}\,\frac{x_R\,K}{\nu_B}\,\gamma_{\mathrm{cool}}^{\alpha}\,,
\end{equation}
where we used Eq.~(7) of \cite{T11} and Eq.~(20) of \cite{B10} and where $g_{\kappa}(p)$ is defined in Eq.~(4) of \cite{T11}.
In the special case B1, we actually first combine Eq.~(\ref{eq1}) and Eq.~(\ref{g_abs}) to get an expression of $\gamma_{\mathrm{cool}}$ independent of $\gamma_{\mathrm{abs}}$. We can now discriminate between the various cases. If $\gamma_{\mathrm{abs}}>\gamma_{\mathrm{cool}}$ occurs we calculate $\alpha$ for the relevant case (A2, B2 or C2) and then update the values of $\gamma_{\mathrm{cool}}$ with Eq.~(\ref{eq2}) and of $\gamma_{\mathrm{abs}}$ with Eq.~(\ref{g_abs}).

We note that the three different cases A, B and C concern the inverse-Compton spectrum and are distinct from the three-stage evolution of the synchrotron self-absorption turnover in the shock model of Marscher \& Gear \cite{MG85}. The latter is governed by the width $x$ of the emission region that can be smaller than $x_R$ if synchrotron or inverse-Compton cooling timescales, $t_{\mathrm{cool}}$, limit the effective emission region to $x_{\mathrm{abs}}=\beta_{\mathrm{rel}}ct_{\mathrm{cool}}(\gamma_{\mathrm{abs}})<x_R$. In the MSSC scenario, $x_{\mathrm{abs}}=x_R(\gamma_{\mathrm{cool}}/\gamma_{\mathrm{abs}})^{\alpha}$ and this inverse-Compton cooling dominates over first-order Compton or synchrotron cooling \cite{BA00}. When $\alpha>1$, $x_{\mathrm{abs}}=x_R(\gamma_{\mathrm{cool}}/\gamma_{\mathrm{abs}})$, where $\gamma_{\mathrm{cool}}$ and $\gamma_{\mathrm{abs}}$ are now controlled by $U_B$ alone (i.e. with $U_S=0$ in Eq.~(8) of \cite{T11}). This synchrotron-cooling stage only sets in if $\gamma_{\mathrm{cool}}<\gamma_{\mathrm{abs}}$ is still satisfied when $\alpha=1$. Otherwise, we have a direct transition from the inverse-Compton stage to the adiabatic expansion cooling stage, where radiative cooling is negligible (i.e. $x_{\mathrm{abs}}>x_R$) and thus the emission region is limited by $x_R$.

\section{Method}
\label{method}
In order to test the MSSC scenario for a shock wave propagating in a relativistic jet flow we had to fully describe the expected evolution with time of both the synchrotron and the inverse-Compton spectra. For this we basically took the same assumptions as described in Sect.~3 of \cite{T11} to define the parameters relevant for the shock-in-jet modeling. We ended-up with an evolution of several key parameters with distance, $X$, of the shock-wave from the apex of the jet, which is itself related to observed time. Those parameters were the normalization, $K$, and the two limits, $\gamma_{\mathrm{min}}$ and $\gamma_{\mathrm{max}}$, of the electron energy distribution, as well as the strength of the magnetic field, $B$, and the maximal thickness of the emission region, $x_R$, which is directly proportional to the jet radius $R$ (cf. Sect.~\ref{theory}). We then calculated the evolution of $\tau_0$ with Eq.~(\ref{tau0}) and derived also that of $\alpha$, $\gamma_{\mathrm{cool}}$ and $\gamma_{\mathrm{abs}}$, which are all smoothly changing with time across the transitions from one case to the other.

We then calculated the evolution of the self-absorption flux density $F_{\mathrm{abs}}\equiv F_{\nu}^{\mathrm{thin}}(\nu_{\mathrm{abs}})$ with Eq.~(6) of \cite{T11}, which normalizes both the synchrotron and inverse-Compton spectra. The exact shape of these components is controlled by $p$ and $\alpha$ for what concerns the spectral slopes and by  $\gamma_{\mathrm{min}}$, $\gamma_{\mathrm{cool}}$, $\gamma_{\mathrm{abs}}$ for the position of the various breaks.
Depending on the relative order of these characteristic Lorentz factors, the synchrotron spectrum can take five different shapes \cite{GS02}, whereas the inverse-Compton spectrum has basically three different shapes for cases A, B and C \cite{B10}.  The analytical formulation used for the breaks has been simplified as described in \cite{T11}. A high-energy cut-off is applied to both spectra at the frequency corresponding to $\gamma_{\mathrm{max}}$.
The inverse-Compton spectrum is also limited at low-frequency by a break towards a spectral index ($F_{\nu}\propto \nu^{-s}$) of $s\!=\!(p-1)/2$ below the rest-frame frequency $\nu_B\gamma_{\mathrm{cool}}^4$ and another one to an index of $s\!=\!-1$ below $\nu_B\gamma_{\mathrm{min}}^4$, or $\nu_B\gamma_{\mathrm{abs}}^4$ if $\gamma_{\mathrm{abs}}>\gamma_{\mathrm{min}}$. Finally, we attempted for a smooth transition from a multiple-Compton to a first-order Compton spectrum.

\begin{figure}[ht]
\begin{minipage}{0.49\textwidth}
\includegraphics[width=\textwidth]{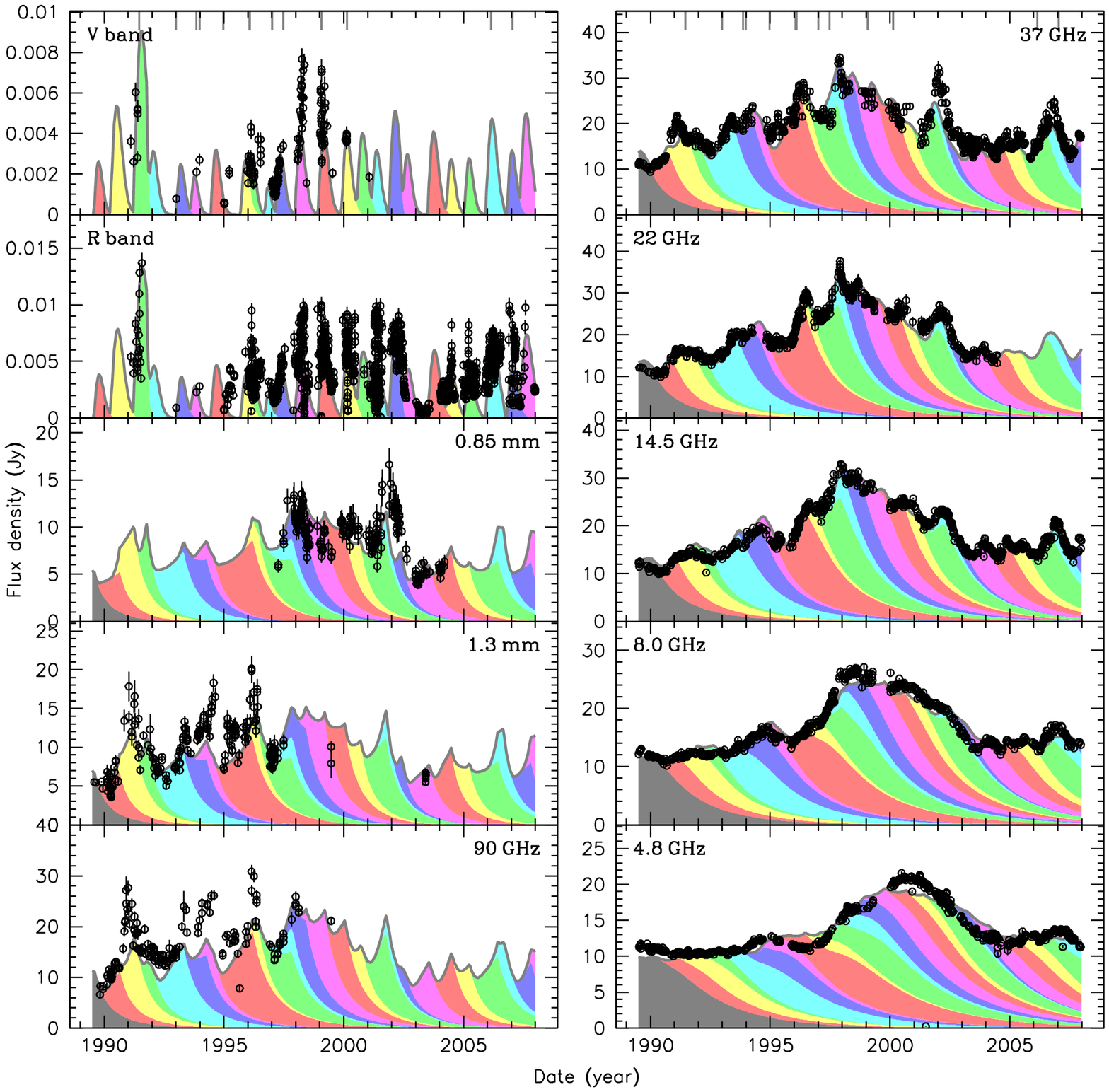}
\caption{\label{fig1}Fit of ten lightcurves of 3C~279 in the optical-to-radio domain with a series of self-similar synchrotron model outbursts following the spectral evolution shown in Fig.~\ref{fig2}. The contribution of individual outbursts is shown with different colors and vertical lines at the top show the times of the available SEDs.}
\end{minipage}\hspace{0.02\textwidth}%
\begin{minipage}{0.49\textwidth}
\includegraphics[width=\textwidth]{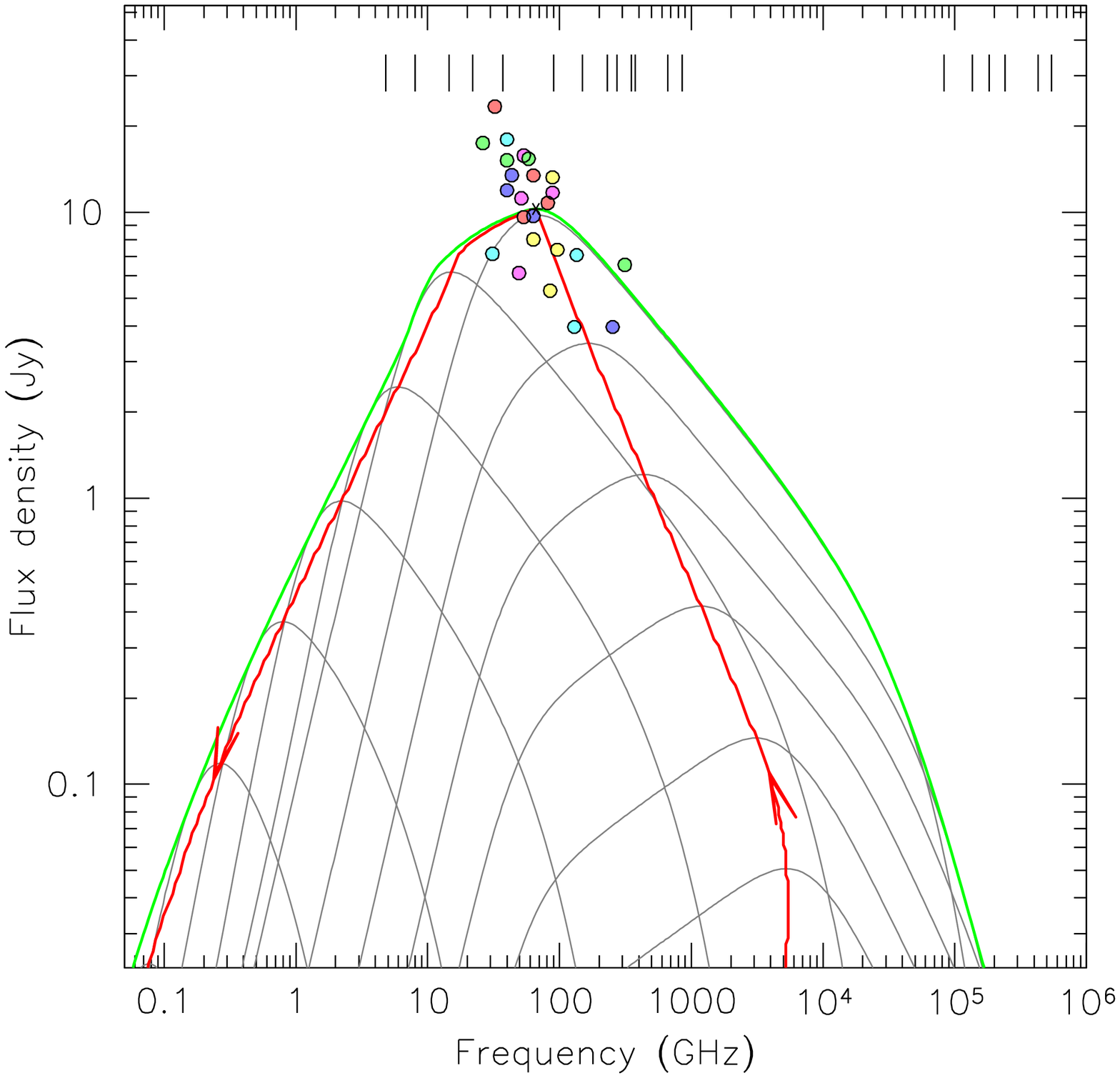}
\caption{\label{fig2}Spectral evolution of the average synchrotron outburst in 3C~279 as derived from the fit of 19 lightcurves (vertical lines) as shown in  Fig.~\ref{fig1}. Synchrotron spectra at different times (grey curves) rise and decay with a maximum following the red line. The peak of individual outbursts is shown by colored dots.}
\end{minipage} 
\end{figure}

\begin{figure}[ht]
\includegraphics[width=0.49\textwidth]{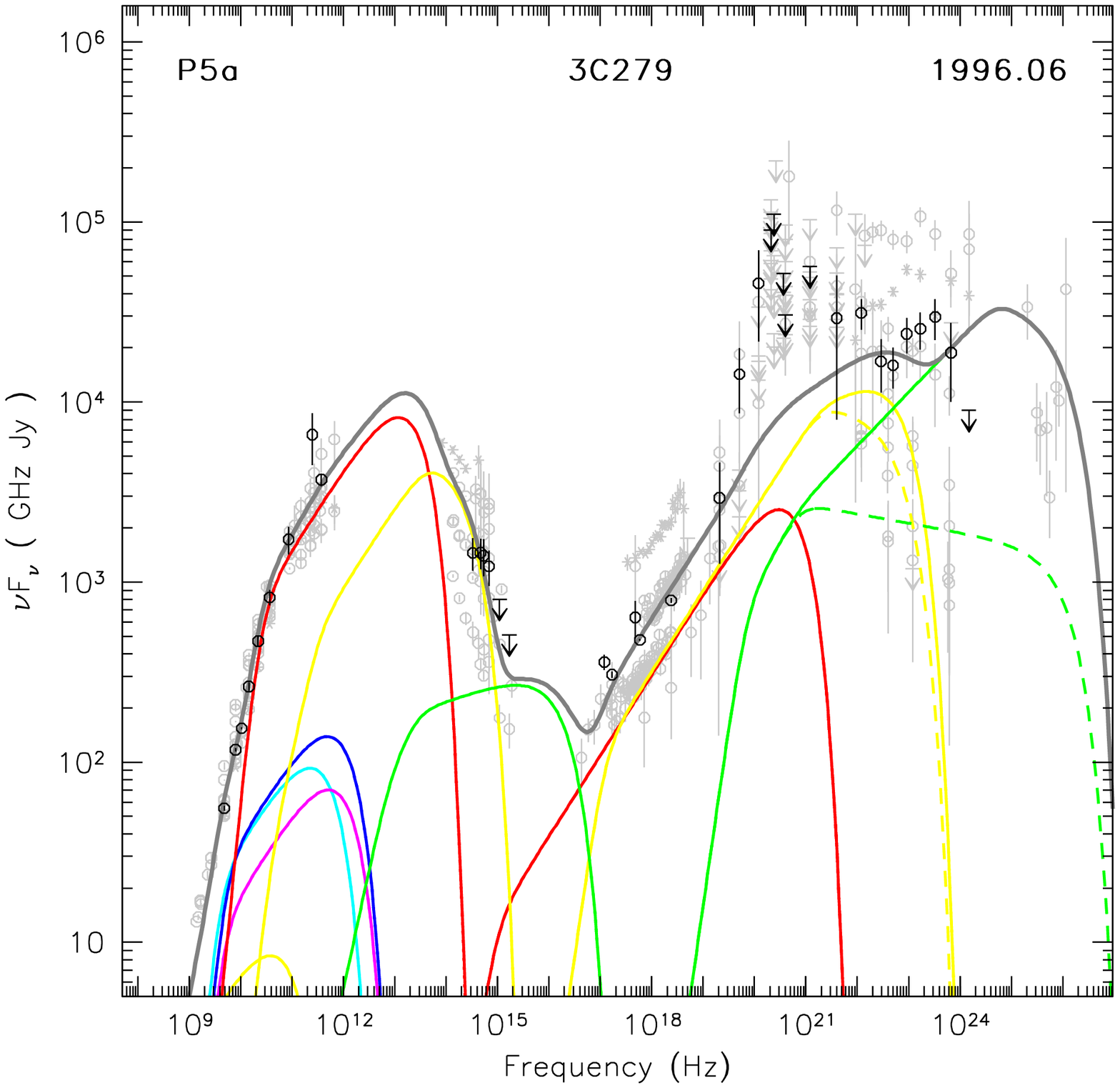}
\hspace{0.02\textwidth}%
\includegraphics[width=0.49\textwidth]{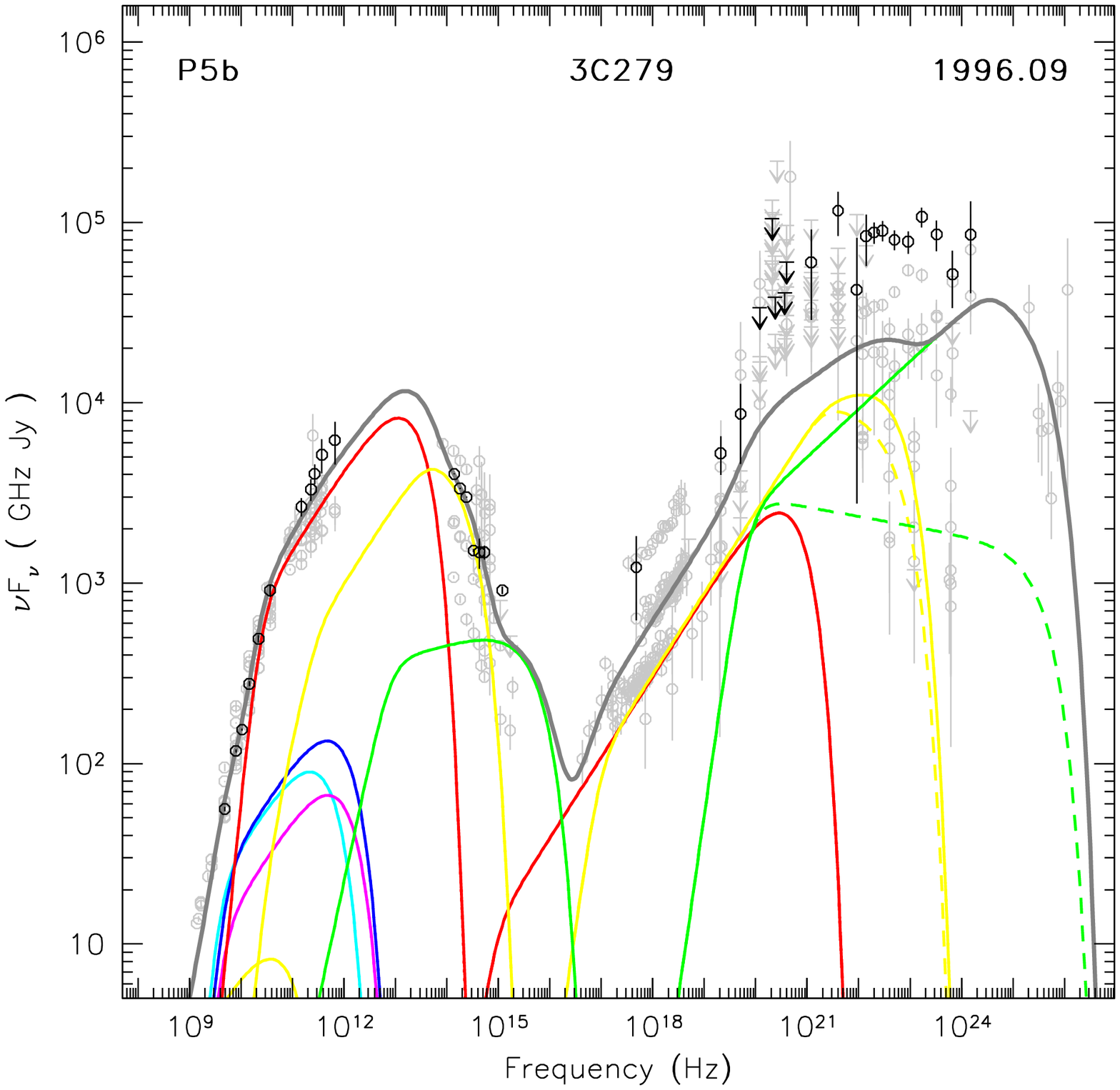}
\caption{\label{fig3}Spectral decomposition of two SEDs of 3C~279 derived by fitting them self-consistently together with all the optical-to-radio lightcurves (see Fig.~\ref{fig1}). The emission of different model outbursts -- corresponding to a succession of different shock waves in the jet -- are shown with the same colors as in Figs.~\ref{fig1} and \ref{fig2}. The important GeV variability between epoch P5b (right panel) and epoch P5a (left panel) in only $\sim$10 days is not fully accounted for by the early emission of a rising outburst (green curves), where MIC scattering dominates strongly over first-order Compton (dashed line). The points from the ten other SEDs and the MAGIC data of 2006 and 2007 \cite{AAA11} are shown in grey for comparison.}
\end{figure}

\section{Results}
\label{results}
To test whether the model outlined above for the high-energy emission of blazars is valid for a flat spectrum radio quasar (FSRQ) like 3C~279, we took all the spectral energy distributions (SEDs) of this object obtained during the operations of the Compton Gamma-Ray Observatory \cite{HBA01}. We then fitted these eleven SEDs together with 19 lightcurves in the radio-to-infrared domain\footnote{The description of the dataset is postponed to another paper in preparation.} with a set of 24 model outbursts corresponding to a succession of shock waves propagating in the jet. Individual outbursts only differ in shock compression factor, $\eta$, and distance from the apex of the jet, $X_{\mathrm{p}}$, needed to build-up the shock \cite{T11}. The obtained fit to the lightcurves is shown in Fig.~\ref{fig1}, while the corresponding synchrotron spectral evolution is shown in Fig.~\ref{fig2}. We note that the latter has a similar shape to what was obtained previously with only about half of the current dataset \cite{LTV06}, but here the rising and slowly decaying peak of the evolution is controlled by a relatively high value of $\gamma_{\mathrm{min}}$ and with no synchrotron stage. The overall spectral decomposition is illustrated in Fig.~\ref{fig3} with the most extreme high-energy spectrum in the dataset (epoch P5b) and the one taken just $\sim$10~days before. In contrary to an early attempt \cite{LVT05}, a pure SSC model seems now to be able to account both for the X-ray and high-energy flaring emission of  3C~279. We note, however, that the presented model is not a true one-zone model. Indeed, to have a deep enough gap between the synchrotron and the inverse-Compton humps, the latter had to be scaled down by a factor 0.1 that corresponds to assuming that the high-energy emission site is 10 times smaller than the synchrotron emission region \cite{B10}.

The model we used here assumes a conical jet with a constant Doppler factor, an opening radius of $2^{\circ}$ pointing $3^{\circ}$ away from the line of sight and with a bulk Lorentz factor for the shocks of $\Gamma\!=\!6.8$. The main parameters for an average outburst at synchrotron peak flux are: a distance $X_{\mathrm{p}}\!=\!16.7$\,pc from the apex of the jet; a magnetic field of $B\!=\!48$\,mG; an electron energy distribution characterized by an index $p\!=\!2.10$, a normalization $K_{\mathrm{p}}\!=\!6460$\,cm$^{-3}$, and cuts at $\gamma_{\mathrm{min}}\!=\!254$ and $\gamma_{\mathrm{max}}\!=\!7450$. This leads to an excess by a factor of $\sim 100$ of the electron energy density $U_{\mathrm{e}}$ compared to $U_B$. Although these values are rapidly changing further upstream or downstream, they seem reasonable compared to other studies (e.g. \cite{HBA01}). The important difference is that our emission site is far more distant from the central engine, where the contribution from photons external to the jet becomes negligible. This is at least true for the long-term X- and gamma-ray emission. The very early evolution of the outburst in our model (see the green spectrum in Fig.~\ref{fig3}) seems to produce flaring emission on time-scales of days at GeV energies and up to to the TeV domain as detected by MAGIC on a few occasions \cite{AAA11}. 

\section{Conclusion}
\label{conclusion}
We presented a new approach to model the variable high-energy emission of blazars and demonstrated its viability by applying it to a very rich dataset of 3C~279 spanning almost 20 years and the full SED from the radio to VHE gamma-rays. These preliminary results suggest that a shock-in-jet model with a proper treatment of MIC scattering can account for the high-energy emission observed in the source at some epochs, while simultaneously reproducing the frequency-dependent variability monitored in the radio-to-optical range.
Such a pure-SSC scenario for FSRQs shall be further tested in a forthcoming paper with the inclusion of the underlying, steady jet emission that shall not exceed the observed quiescent state of the source and by including addition observational constraints in the X-ray to gamma-ray range, in particular with lightcurves of 3C~279 from the satellites  \emph{Fermi},  \emph{Swift}, and the \emph{Rossi X-ray Timing Explorer} (e.g. \cite{AAA10}).

\ack{\small This work was done in the frame of the International Team collaboration number 160 supported by the International Space Science Institute in Bern, Switzerland.}

{\footnotesize
\bibliography{turler}	
}

\end{document}